\documentclass{aa}  

\usepackage{natbib}
\bibpunct{(}{)}{;}{a}{}{,}  
\usepackage{amsmath}
\usepackage{graphicx}
\usepackage{longtable}
\usepackage[varg]{txfonts}
\begin{document}
   \title{The nature of the KFR08 stellar stream
   		\thanks{Based on observations made with Nordic Optical Telescope at La Palma under programme 44-014 and on data obtained from the ESO Science Archive Facility under programme ID 071.B-0529(A), 072.C-0488(E), 077.C-0192(A), 082.B-0610(A), 085.C-0062(A).}}

   \subtitle{A chemical tagging experiment}

   \author{C. Liu
   	\inst{1}
          \and
          S. Feltzing\inst{1}
          \and
         G. Ruchti\inst{1}
          }

   \institute{Lund Observatory, Department of Astronomy and Theoretical Physics,
              Box 43, SE--221 00 Lund, Sweden\\
              \email{[cheng, sofia, greg]@astro.lu.se}
              }

   \date{Received 23 December 2014; accepted 24 June 2015}

 
  \abstract
{The origin of a new kinematically identified metal-poor stellar stream, the KFR08 stream, has not been established. We present stellar parameters, stellar ages, and detailed elemental abundances for Na, Mg, Al, Si, Ca, Sc, Ti, Cr, Ni, Zn, Sr, Y, Zr, Ba, La, and Eu for 16 KFR08 stream members based on analysis of high resolution spectra. Based on the abundance ratios of 14 elements, we use the chemical tagging method to identify the stars which have the same chemical composition, and thus, might have a common birthplace, such as a cluster. Although three stars were tagged with similar elemental abundances ratios, we find that, statistically, it is not certain that they originate from a dissolved star cluster. This conclusion is consistent with the large dispersion of [Fe/H] ($\sigma_{\rm{[Fe/H]}} = 0.29$) among the 16 stream members. We find that our stars are $\alpha$ enhanced and that the abundance patterns of the stream members are well matched to the thick disk. In addition, most of the stream stars have estimated stellar ages larger than 11 Gyr. These results, together with the hot kinematics of the stream stars, suggest that the KFR08 stream is originated from the thick disk population which  was perturbed by a massive merger in the early universe.}


   \maketitle
%

\section{Introduction}

Star streams, which are groups of stars on the same orbit in the Galactic potential, have been detected as over-densities in the velocity distribution of stars in the solar neighbourhood (e.g., \citealp{1998AJ....115.2384D, 2005A&A...430..165F, 2006A&A...449..533A}). \cite{1999Natur.402...53H} discovered the signature of a cold stream in the velocity distribution of the halo and interpreted this stream as part of the tidal debris of a disrupted satellite galaxy accreted by the Milky Way. Moreover, several studies (\citealt{2004ApJ...601L..43N, 2006MNRAS.365.1309H, 2006A&A...449..533A}) have concluded that the Arcturus stream is another such debris stream dating back to an accretion event 5 to 8 Gyr ago. The external origin of such halo streams is supported by numerical simulations \citep{2003ApJ...592L..25H} and observations of ongoing satellites accretion such as that of the Sagittarius dwarf galaxy \citep{1994Natur.370..194I}.

However, accreted satellites are not the only source of streams. A stellar stream that is homogenous in age and chemical composition is associated with a dissolved star cluster. One striking example is the HR 1614 stream \citep{1978ApJ...222..191E, 2000A&A...357..153F, 2007AJ....133..694D}. Stellar streams can also originate from dynamical effects within the disks due to resonances with the bars or spiral arms \citep{2000AJ....119..800D, 2009ApJ...700L..78A}. Analysis of high-resolution spectra of nearby F and G dwarf stars has revealed that the stars in the Hercules stream have a wide range of stellar ages, metallicities and element abundances \citep{2007ApJ...655L..89B}. They concluded that the kinematical properties of the Hercules stream are coupled to dynamical interactions with the Galactic bar.

Recently, a new candidate stream, called the KFR08 stream, on a quite radial orbit was discovered by \cite{2008ApJ...685..261K} exploring the RAVE DR1 experimental data \citep{2006AJ....132.1645S}. This stream is present as a broad feature in the range $-180~\mathrm{km~s^{-1}} \le V \le -140 ~\mathrm{km~s^{-1}}$ centered at $V \approx-160~\mathrm{km~s^{-1}}$ and would belong to the stellar halo population. The velocities, including $U$ and $W$, here are given relative to the local standard of rest (LSR). Because of the high $W-$velocities of the stars in the stream, an origin external to the Milky Way's disk was suggested by \cite{2008ApJ...685..261K}. This is supported by \cite{2009ApJ...698..865K} and \cite{2010AstL...36...27B} analyzing two independent dat sets. \cite{2009ApJ...698..865K} characterized the orbits of calibrated stars from SDSS DR7 data \citep{2009ApJS..182..543A} through angular momentum, eccentricity, and orbital polar angle. An overdensity region that corresponds to the location of the KFR08 stream within parameter distributions was found. Based on a new version of the Hipparcos catalogue \citep{2007A&A...474..653V} and updated Geneva-Copenhagen Survey \citep{2007A&A...475..519H} of F and G dwarfs, \cite{2010AstL...36...27B} identified statistically significant signals of the main inhomogeneities in the velocity distribution using wavelet transform technique. They found that 19 stars cluster around (--160, 225)~$\mathrm{km~s^{-1}}$ in the $V$ and $(U^{2}+2V^{2})^{1/2}$ plane. However, the KFR08 stream could not be completely confirmed by re-analyzing the pure RAVE DR2 dwarf sample \citep{2011ApJ...726..103K}. Five stars, at the phase-space position of the KFR08 stream, identified from DR2 have a large scatter in metallicity. This is inconsistent with the prediction of chemical homogeneity of stream members if they come from a single cluster. But it might agree with the idea proposed by \cite{2009MNRAS.396L..56M} that this high-velocity stream has a dynamical origin in the thick disc. It is thought to arise due to a sudden energy kick imposed by a massive satellite in the past. The signature of this perturbation can be identified in the stellar kinematics if it was not wiped out by the radial mixing \citep[see][]{2010ApJ...713..166B} within the thick disc.

Although brilliant works on dynamical analysis have been done, the detailed elemental abundance and age structure of the KFR08 stream have never been investigated. Analysis of high-resolution spectra of possible stream members could give us more clues for exploring the origin of the stream. In this work, we present how candidates of the stream were selected from databases and how the spectra of them were observed using different instruments in Section 2. Stellar parameters, ages and elemental abundances are determined from the spectra in Section 3. In Section 4, we use chemical tagging to identify the putative cluster stars from kinematically detected KFR08 stream. The possible origins of the stream are discussed in Section 5. Finally, conclusions are drawn in Section 6.


\section{Observations}

\subsection{The stellar sample}

\cite{2010AstL...36...27B} found nineteen stars that are likely members of the KFR08 stream. Of those 17 are high probably members. For 16 of these we have analyzed high resolution spectra. These stars and their properties are listed in Table~\ref{sp}. We also give their Galactic coordinates, distances to the Sun and space velocities with respect to the LSR in Table~\ref{age}.

\subsection{Spectroscopic observations}

\paragraph{FIES observation:} 
Observations were carried out at the Nordic Optical Telescope (NOT) for 10 of the candidates using the fibre-fed Echelle Spectrograph (FIES) on January 10 -- 12 in 2012. A solar spectrum was also obtained by observing the sky at daytime. The wavelength range of the spectra is 370 -- 730 nm, with a resolving power R $\sim$ 67000 and the average signal-to-noise ratio (S/N) is larger than 100 per pixel for most of the spectra. All the spectra were reduced using the FIEStool\footnote{http://www.not.iac.es/instruments/fies/fiestool/FIEStool.html} pipeline. The pipeline includes the following steps to reduce the observed frame: subtracting bias and scattered light, dividing by a normalized two-dimensional flat field, extracting individual orders, and solving wavelength. Finally, all individual spectral orders are merged into a one-dimensional spectrum. 

\paragraph{Archival data:}
Reduced 1D spectra for 6 stars were extracted from the ESO archive. We found that 3 stars have been observed using the UVES spectrograph \citep{2000SPIE.4008..534D} on the VLT 8-m telescope in 2013. Two spectra have medium resolution R $\sim$ 45000, while the spectrum of star HIP 59785 was obtained in high resolution mode (R $\sim$ 110000). The S/N values of those three spectra are larger than 100 per rebined pixel. The spectrum for HIP 117702 was observed in 2006 with FEROS on ESO 2.2m-telescope \citep{1999Msngr..95....8K}. This spectrum has a large wavelength range (350--920 nm), high resolution (R $\sim$ 48000) and good S/N ($\sim$ 75 per rebined pixel). The exoplanet survey carried out with HARPS on nearby stars with high resolution spectroscopy (R $\sim$ 120000, \citealt{2003Msngr.114...20M}) includes spectra for two of the stars identified as possible stream members in \cite{2010AstL...36...27B}.  Both spectra have a mean S/N larger than 50 per rebined pixel. 

Radial velocities (RVs) were measured by cross-correlating the solar synthesis spectrum with the observed spectra, making use of the packages $\mathit{RVSAO}$ and $\mathit{XCSAO}$ within IRAF \footnote{IRAF is distributed by the National Optical Astronomy Observatory, which is operated by the Association of Universities for Research in Astronomy (AURA) under cooperative agreement with the National Science Fundation.}. Both our own spectra and collected spectra from the archive are shifted to rest wavelength by cross-correlation with a synthetic solar spectrum using the IRAF task $\mathit{DOPCO}$. 


\section{Abundances Analysis}

\subsection{Stellar Parameters}

For our spectral analysis, we use Spectroscopy Made Easy (SME, \citealt{1996A&AS..118..595V, 2005ApJS..159..141V}) to determine both stellar parameters and elemental abundances by comparing synthetic spectra with observations. In SME the model atmospheres are interpolated in the precomputed MARCS model atmosphere grid \citep{2008A&A...486..951G}, which have standard composition. 

\begin{figure}[htbp]
\begin{center}
\includegraphics[scale=0.5]{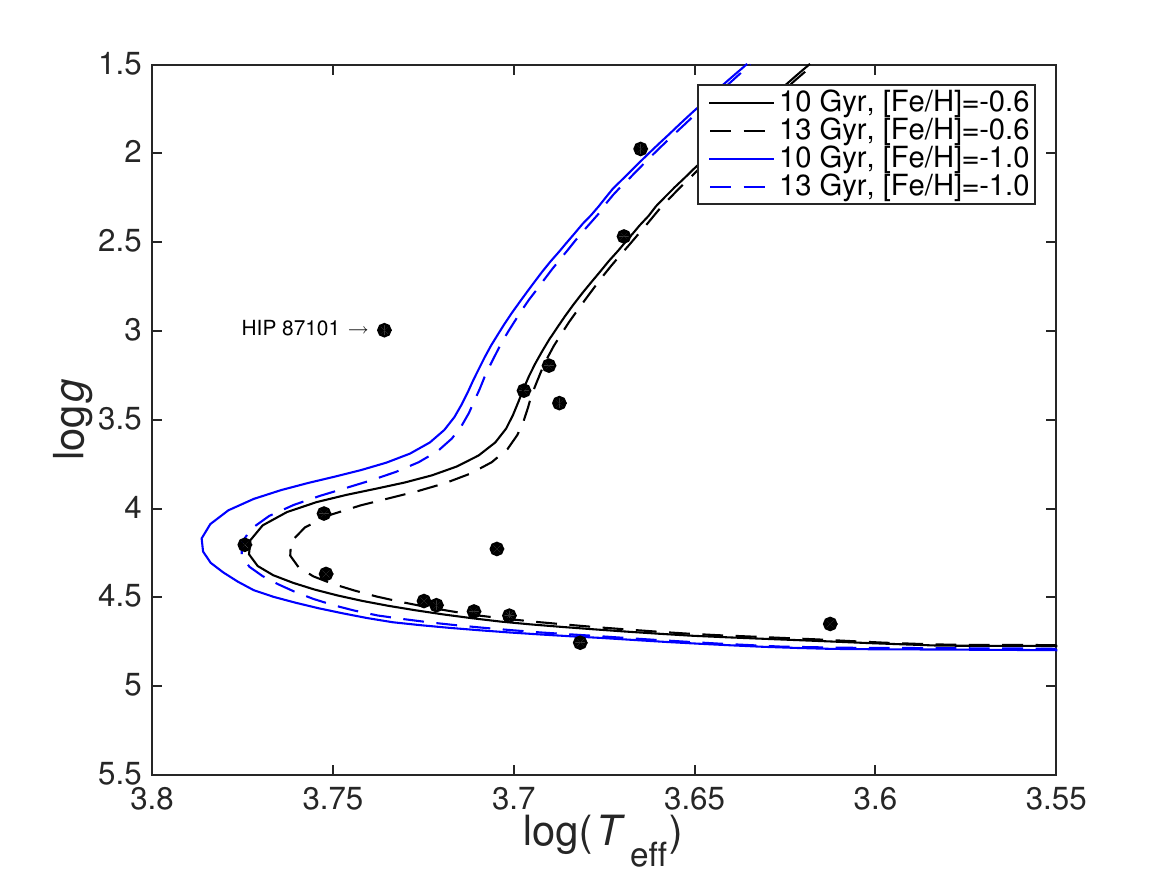}
\caption{$T_{\rm eff}$ vs. $\log g$ diagram. Filled circles indicate our main sample from Hipparcos catalogue. Isochrones at different ages (10 and 13 Gyr) and metallicities (--0.6 and --1.0 dex) according to the Yonsei-Yale models \citep{2004ApJS..155..667D} are shown.}
\label{cmd}
\end{center}
\end{figure}

The stellar parameters ($T_{\rm eff}$, $\log g$, [Fe/H]) were determined following the same methodology as in \citet[hereafter L15]{2015A&A...575A..51L}. We refer the reader to L15 for full details. In brief, initial stellar parameters which are used to generate a synthetic spectrum were estimated before running SME. As most of the candidates have metallicities around --0.7 dex \citep[Table 2 in][]{2010AstL...36...27B}, we assume an initial [Fe/H] of --0.7 dex for all stars. Following the same methods as in L15, initial effective temperatures ($T_{\rm eff}$) were determined using the colour-metallicity-temperature relations of \cite{1996A&A...313..873A}, using both $B-V$ and $b-y$ colors, where $B-V$ came from the Hipparcos catalogue \citep{1997A&A...323L..49P} and $b-y$ values were obtained from the Str\"{o}mgren survey \citep{1983A&AS...54...55O, 1988A&AS...73..225S, 1993A&AS..102...89O}. An initial estimate of surface gravities ($\log g$) were calculated using $T_{\rm eff}$, distances (from Hipparcos parallaxes), bolometric corrections, and stellar mass from the Yonsei-Yale isochrones \citep{2004ApJS..155..667D}. Then, the stellar parameters (shown in Table~\ref{sp}) were determined by using a purely spectroscopic method (called Procedure 1 in L15). The location of stars in $T_{\rm eff}$ vs. $\log g$ diagram are shown in Fig~\ref{cmd}. Considering systematic errors and possible sources of uncertainty in the atmospheric model and atomic line data, the systematic uncertainties in the stellar parameters were estimated to be $\delta T_{\rm eff}=67 \pm 40$ K, $\delta {\rm log}$ $g = 0.08 \pm 0.06$ dex, and $\delta \rm{[Fe/H]} = 0.06 \pm 0.03$ dex.

Since our stellar parameters are derived using Fe lines, which were compiled for the spectra of solar type stars, it is possible that the results may include systematic uncertainties due to departures from non-local thermodynamical equilibrium \citep[NLTE, e.g.][]{2012MNRAS.427...27B}. Following the same methodology as in \cite{2013MNRAS.429..126R}, we derived stellar parameters for two stars (HIP 5336 and HIP 58708) using on the fly NLTE corrections for Fe. It was found that the mean differences between our results and stellar parameters determined from equivalent width method are $\Delta T_{\rm eff}= -67$ K, $\Delta {\rm log}~g = 0.02$ dex and $\Delta$ [Fe/H] = --0.02 dex, respectively. These differences are well within our estimated uncertainties, and will not effect our chemical tagging experiment.

As the precision of the parallax for all stars is better than 15\%, based on the partly physical method (called Procedure 2 in L15), we also measured their stellar parameters (Table~\ref{sp}) by fixing $\log g$ as estimated from the Hipparcos parallaxes. Comparison with the $\log g$ estimated from the Fe lines, the differences between the two surface gravities are less than 0.15 dex for most of the stars. An exception is found for two stars, HIP 59785 and HIP 87101, their $\log g$ are changed by 0.6 dex. It was also found that the mean differences in $T_{\rm eff}$ are larger than 200 K for those two stars. Since HIP 59785 is a very cool star with $B-V> 1.0$, we might overestimate $\log g$ from parallax caused by the wrong bolometric correction. Because this cool star is close to the colour limits for the bolometric correction (see \citealt{2010AJ....140.1158T}). For HIP 87101, it is apparent that the location of this star is far from the isochrones in the Fig.~\ref{cmd}. However, the $T_{\rm eff}$ and $\log g$ based on the parallaxes are coupled with the isochrones. It should also be notice that HIP 87101 is the most metal-poor star in our sample. It might be subject to strong NLTE effects \citep[see][]{2012MNRAS.427...27B}. These imply that we might obtain unreliable stellar parameters from the Procedure 1 for this star. We noticed that the difference in [Fe/H] is --0.28 dex for the star HIP 55988. The noised Fe lines might induce this huge difference in [Fe/H] because of the lowest S/N ($<$ 30 per rebined pixel) spectrum among the stars. Except for these three stars, the mean difference between two obtained stellar parameters are $\Delta T_{\rm eff}=59 \pm 66$ K, $\Delta \log g = 0.12 \pm 0.16$ dex, and $\Delta \rm{[Fe/H]} = 0.03 \pm 0.04$ dex,  respectively. Since a little changes in [Fe/H] and $T_{\rm eff}$ do not affect our final results from the chemical tagging experiments (see Sect.~4). We can conclud that two previous derived stellar parameters are consistent at certain level. For the remainder of our analysis we adopt the stellar parameters derived using a purely spectroscopic method.

\subsection {Stellar Ages}

Using the stellar parameters derived from the spectra, stellar ages were estimated from fits to the Yonsei-Yale isochrones \citep{2004ApJS..155..667D} by maximising the probability distribution functions as described in \cite{2011A&A...533A.134B}. The most probable age is determined from the peak of the age probability distribution, 1$\sigma$ lower and upper age limits are obtained from the shape of the distribution. Stellar masses were also determined in a similar manner. Both the ages and masses are listed in Table~\ref{age}. It is clearly seen that most of the stars, for which ages can be determined, have ages larger than 11 Gyr. The possible systematic biases mainly caused by sampling the isochrone data points \citep{2004A&A...418..989N} in our probabilistic age determinations was discussed in L15. Comparing with the given typical errors in ages, they found that it can be ignored.

\begin{table*}[htbp]
\centering
\setlength{\tabcolsep}{6pt}
\begin{tiny}
\caption{Stellar parameters of KFR08 stream members.}
\label{sp}
\begin{tabular}{lccrrrrrcccrrrr}
      \hline
      \hline
      Names  &$T_{\rm eff}$  &log $g$  &[Fe/H]  &$v_{mic}$ &$v_{mac}$ &$v{\rm sin}i$ &$T'_{\rm eff}$  &$\log g'$  &[Fe/H]$'$  &$v'_{mic}$ &$v'_{mac}$ &$v{\rm sin}i'$\\
        HIP          &(K)     &    &    &(km s$^{-1}$)     &(km s$^{-1}$) &(km s$^{-1}$) &(K)     &    &    &(km s$^{-1}$)     &(km s$^{-1}$) &(km s$^{-1}$)\\
      \hline
      5336            &5304  &4.52  &--0.77   &1.0  &2.0  &0.7  &5354  &4.60  &--0.76   &0.7   &2.0   &0.6\\
   15495            &5029  &4.60  &--0.57   &0.6  &1.0  &0.6  &5119  &4.79  &--0.55   &0.7   &0.7   &0.8\\
   18235            &4980  &3.33  &--0.68   &0.6  &3.2  &1.0  &4976  &3.30  &--0.71   &1.1   &3.3   &0.7\\
   19143            &4803  &4.76  &--0.88   &1.0  &0.9  &0.5  &4939  &5.07  &--0.84   &0.5   &0.3   &0.4\\
   54469            &5948  &4.20  &--0.45   &1.0  &3.0  &2.7  &6093  &4.42  &--0.42   &1.2   &1.8   &3.5\\
   55988            &4096  &4.65  &--0.82   &0.4  &--0.1  &0.1 &4071  &5.08  &--0.54  &0.2  &0.0   &0.1\\
   58357            &4898  &3.20  &--0.67   &0.6  &3.3  &0.9   &4912  &3.25  &--0.68   &1.1   &3.4  &0.8\\
   58708            &4869  &3.41  &--0.60   &1.0  &3.0  &0.9   &4924  &3.37  &--0.56   &1.1   &3.1  &0.8\\
   58843            &5649  &4.37  &--0.76   &1.0  &2.9  &0.5  &5636  &4.39  &--0.79   &1.0   &2.9   &0.5\\
   59785            &4620  &1.98  &--0.63   &1.5  &4.9  &1.9  &4808  &2.52  &--0.48   &1.7   &4.7   &1.5\\
   60747            &5268  &4.54  &--0.85   &0.5  &1.5  &0.6  &5319  &4.65  &--0.84   &0.6  &1.6  &0.6\\
   64920            &4672  &2.47  &--0.58   &1.2  &0.9  &4.1  &4701  &2.51  &--0.57   &1.2   &0.7   &4.3\\
   74033            &5657  &4.03  &--0.73   &1.1  &3.5  &1.3  &5646  &4.02  &--0.75   &1.1   &3.3   &1.8\\
   81170            &5069  &4.23  &--1.22   &0.8  &0.0  &0.1  &5280  &4.78  &--1.13   &0.5  &--0.2  &0.1\\
   87101            &5485  &3.14  &--1.68   &1.5  &--0.2  &0.1  &5875  &4.23  &--1.43   &1.3  &--0.5  &0.1\\
 117720            &5141  &4.58  &--0.56   &0.6  &0.5  &1.5  &5150  &4.63  &--0.65   &0.8   &1.0   &0.8\\
           \hline
\end{tabular}
\tablefoot{The second to seventh columns give the global parameters obtained through purely spectroscopic approach (Procedure 1, L15), while the eighth to thirteenth columns give the six parameters obtained from Procedure 2 in L15. The unprimed quantities are the preferred values which are used to estimate the stellar masses and ages.}
\end{tiny}
\end{table*}

\begin{table*}[htbp]
\centering
\setlength{\tabcolsep}{4pt}
\begin{tiny}
\caption{The Galactic coordinates, distances, space velocities, radial velocities, masses, absolute magnitudes and ages of KFR08 stream members.}
\label{age}
\begin{tabular}{lrrrrrrrrcccrrrr}
      \hline
      \hline
      Names &$l$ &$b$ &$d$ &$\sigma_{d}$ &U$^{a}$ &V$^{a}$ &W$^{a}$ &V$_{r}$ &$\sigma_{\rm{V}_{r}}$ &M &$M_{V}$ &$\sigma_{M_{V}}$ &Age &-1$\sigma$ &+1$\sigma$\\
        HIP    &(deg) &(deg) &(pc) &(pc) &(km s$^{-1}$) &(km s$^{-1}$) &(km s$^{-1}$) &(km s$^{-1}$) &(km s$^{-1}$) &M$_{\odot}$&(mag) &(mag) &(Gyr) &(Gyr) &(Gyr)\\
      \hline
      5336           &125.3637   &--07.8683   &7.6    &0.1 &--32 &--153 &--28 &--96.2  &0.2 &0.70 &5.76 &0.01 &14.7 &9.9 &---\\
   15495            &155.1775    &--19.8313   &46.2    &2.8 &58 &--174 &--3 &--104.5 &0.2 &0.69 &6.34 &0.13 &13.1 &6.3 &13.5\\
   18235            &217.8451    &--48.7113   &64.2    &2.4 &--16 &--161 &--19 &120.6 &0.5 &0.87 &2.65 &0.08 &14.7 &9.3 &---\\
   19143            &163.4802    &--14.1808   &36.6    &2.3 &--140 &--143 &--42 &109.9 &0.3 &0.61 &7.15 &0.14 &6.4 &3.2 &12.1\\
   54469            &284.1956     &14.8416   &107.9  &15.0 &91 &--159 &--64 &156.4 &0.5 &0.94 &4.66 &0.30 &8.9 &8.1 &10.5\\
   55988            &254.0288     &62.2288   &27.4    &1.3 &50 &--154 &--25  &23.6 &0.4 &0.50 &8.02 &0.10 &14.6 &7.8 &---\\
   58357            &145.6480     &66.4998  &146.0   &21.1 &--123	 &--134 &45 &47.2 &0.1 &0.87 &2.53 &0.31 &14.7 &9.9 &---\\
   58708            &151.6559     &71.2086   &57.1   &1.7 &--14 &--160 &15 &--10.6 &0.1 &0.87 &2.84 &0.06 &14.8 &11.5 &---\\
   58843            &275.4489     &63.6586   &70.2    &6.6 &122 &--138 &--58 &10.3 &0.1 &0.77 &4.99 &0.20 &14.8 &12.7 &---\\
   59785            &295.8082     &20.4609   &123.2   &6.8 &--117 &--136 &--109 &25.2 &1.0 &1.92 &0.8 &0.12 &0.6 &--- &6.5\\
   60747            &288.9989     &63.7756   &81.9   &10.7 &110 &--146 &91 &153.4 &0.2 &0.68 &5.91 &0.28 &14.6 &9.7 &---\\
   64920            &310.1019     &34.1637   &165.6  &15.6 &66 &--159 &43 &140.3 &0.5 &0.88 &0.70 &0.20 &11.0 &5.7 &13.1\\
   74033            &9.8705     &53.0007   &71.9    &6.3 &--113 &--132 &42 &--59.4 &0.2 &0.83 &3.98 &0.19 &14.5 &11.9 &---\\
   81170            &11.6638     &27.7129   &45.1   &2.8 &--77 &--157 &--123 &--170.9 &0.1 & &6.34 &0.13 & & &\\
   87101            &16.9365      &9.4726   &110.0  &15.4 &--76 &--159 &--3 &--129.5 &0.3 &1.22 &4.48 &0.30 &1.9 &1.8 &7.0\\
 117720            &315.0151    &--54.2925   &50.4    &2.7 &12 &--159 &124 &--25.4 &0.9 &0.70 &5.93 &0.12 &13.1 &6.7 &13.6\\
           \hline
\end{tabular}
\tablefoot{"--" indicates the 1$\sigma$ lower (or upper) age could not be determined because a very young (or old) star is out side of the isochrones limitations. $^{a}$ means that the space velocities of the stars come from \cite{2010AstL...36...27B}.}
\end{tiny}
\end{table*}

\subsection{Elemental Abundances}

Abundances of $\alpha$-elements (Mg, Si, Ca, Ti), iron peak elements (Cr, Ni, Zn), odd-Z elements (Na, Al), s-process elements (Sr, Y, Zr, Ba, La), r-process (Eu), and Scandium were measured by fitting the selected absorption lines for each element simultaneously. Except for Ba, La, and Eu, we combined the line lists from the \cite{2003A&A...410..527B} and the Gaia-ESO line list (Heiter et al. in prep) as we did in L15. In this work, the average abundance of Ti and Cr were derived by measuring both neutral and singly ionized lines. Since hyperfine splitting (HFS) has the effect of desaturating strong lines \citep{1995AJ....109.2757M}, we analyzed the four Ba II lines by adopting the HFS from \cite{1998AJ....115.1640M}. And for the La and Eu lines, we adopted the data from \cite{2001ApJ...556..452L} and \cite{2001ApJ...563.1075L}, respectively. In addition, isotopes further contribute to the broadening of Ba, La, and Eu lines \citep[e.g.][]{2001ApJ...563.1075L, 2002ApJ...566L..25S}. We assumed solar system isotopic ratios for these elements.

\begin{figure*}[htbp]
\begin{center}
\includegraphics[scale=0.4]{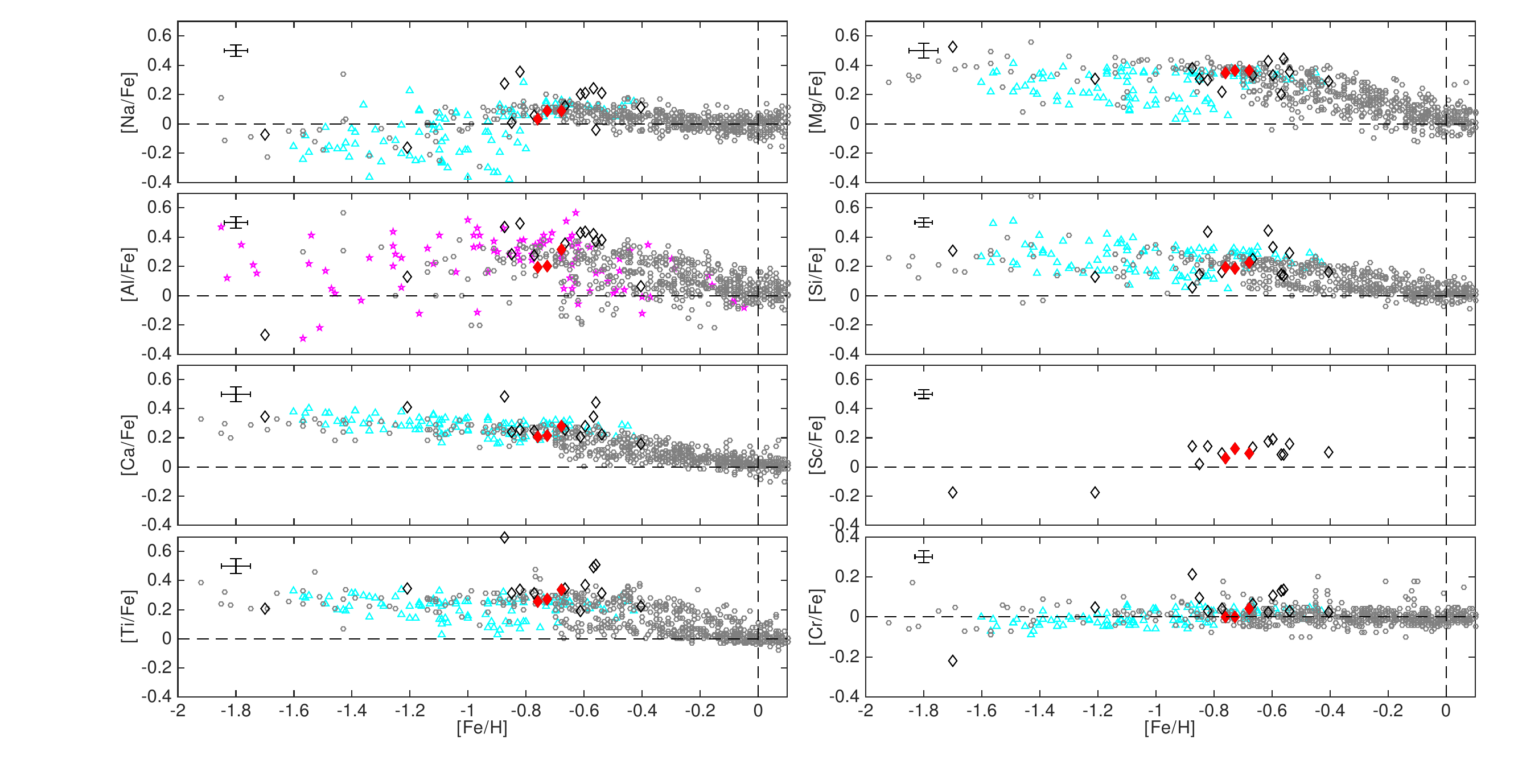}
\caption{Light elemental abundance ratios [X/Fe] as the function of [Fe/H]. Three tagged cluster stars (see Sect.~4) are indicated in red filled diamonds, while other stream stars are marked by black emptied diamonds. The disk stars from \cite{2014A&A...562A..71B} are marked open gray circles, while the cyan triangles and magenta stars indicate the thick disk and halo stars collected from \cite{2010A&A...511L..10N} and \cite{2000AJ....120.1841F}, respectively. Dashed lines indicate solar values.}
\label{abunp1}
\end{center}
\end{figure*}

\begin{figure*}[htbp]
\begin{center}
\includegraphics[scale=0.4]{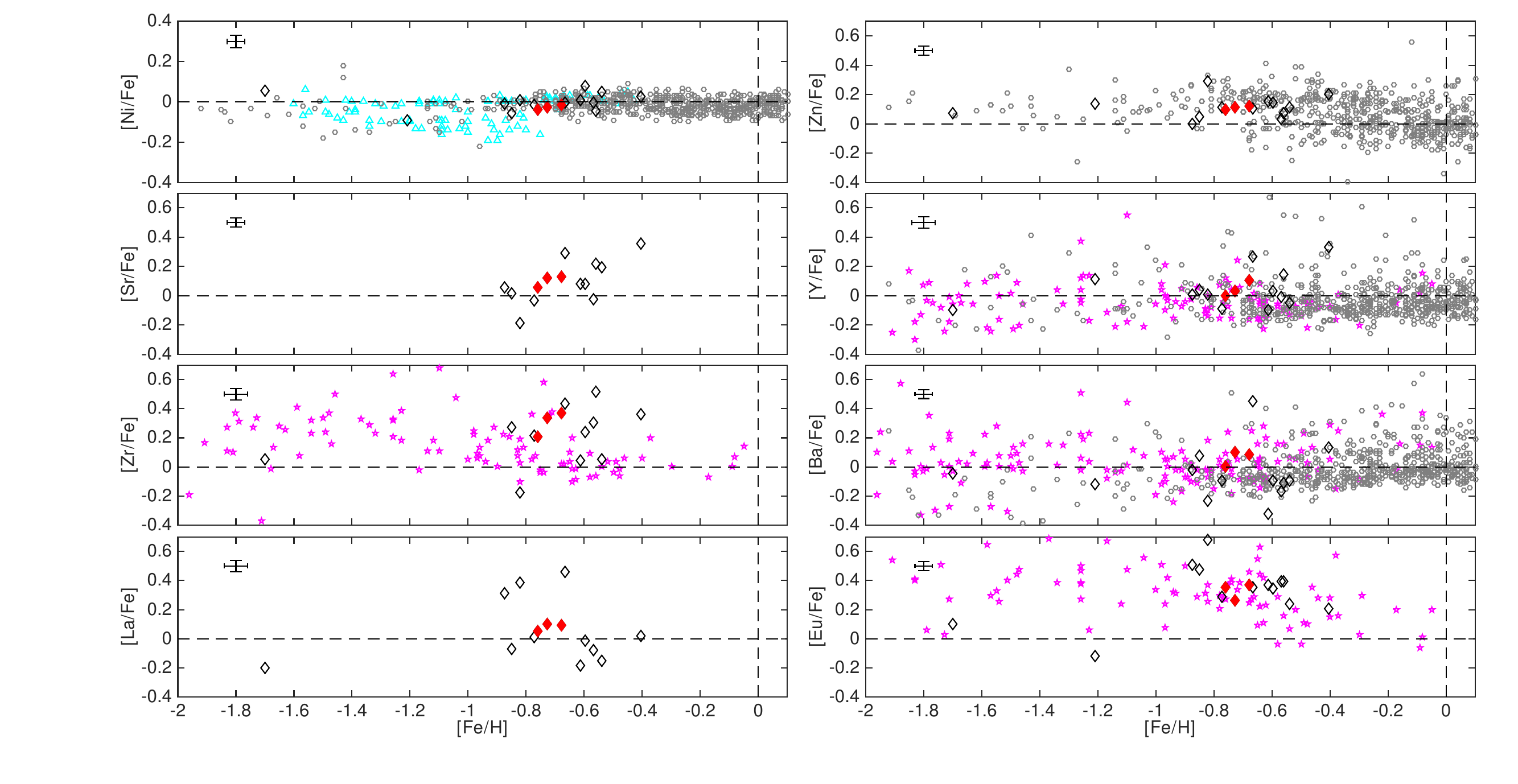}
\caption{Heavy elemental abundance ratios [X/Fe] as the function of [Fe/H]. The symbols and colors have the same meaning as the Fig.~\ref{abunp1}. Dashed lines indicate solar values.}
\label{abunp2}
\end{center}
\end{figure*}

During the abundance analysis, we left the corresponding elemental abundance (e.g. [Na/H]) free while the stellar parameters were kept fixed. The average ratio $A_{\rm{Na}}$ (the absolute abundance relative to the total number density of atoms) is the output from SME. The solar elemental abundance values taken from \cite{2007SSRv..130..105G} were used to normalize the SME results, in order to obtain the abundance ratio, e.g., [Na/H]. The abundance ratios with respect to Fe (i.e. [X/Fe] in standard notation) were also calculated and are shown in Fig.~\ref{abunp1} and~\ref{abunp2} and in Table~\ref{abun}. To normalize the determined abundances of our sample, we also derived solar abundances, using the same line list, and the stellar parameters derived from our solar spectrum in Sect. 3.1. 

\begin{table*}[htbp]
\centering
\setlength{\tabcolsep}{6pt}
\begin{tiny}
\caption{The elemental abundances of KFR08 stream members.}
\label{abun}
\begin{tabular}{lrrrrrrrrrrrrrrrrrr}
      \hline
      \hline
                      &---     &---    &---   &---  &---  &--- &---   &[X/Fe]    &---   &---    &---  &---  &--- &--- &--- &---  \\
      HIP   &Na &Mg &Al &Si &Ca &Sc &Ti &Cr &Ni &Zn &Sr &Y &Zr &Ba &La &Eu \\
               &             &      &       &    &      &     &      &     &     &    &    &    &   &  &  &    \\
      \hline
      5336         &0.05 &0.22 &0.27 &0.16 &0.25 &0.10 &0.31 &0.04 &--0.02 &0.11	 &--0.03 &--0.09 &0.22 &--0.10 &0.01 &0.29 \\
     15495        &0.25 &0.20 &0.42 &0.14 &0.35 &0.09 &0.49 &0.13 &0.00 &0.03 &--0.02 &--0.01 &0.30 &--0.16 &--0.08 &0.40 \\
    18235         &0.09 &0.36 &0.31 &0.23 &0.28 &0.10 &0.34	 &0.04 &--0.01 &0.12 &0.13 &0.11 &0.37 &0.08 &0.09 &0.37 \\
     19143        &0.27 &0.38 &0.47 &0.06 &0.49 &0.14 &0.70 &0.21 &--0.01 &0.00 &0.06 &0.01 &0.75 &--0.02 &0.31 &0.51 \\
     54469        &0.11 &0.29 &0.06 &0.17 &0.16 &0.10 &0.22 &0.02 &0.03 &0.20 &0.36 &0.33 &0.36 &0.14 &0.02 &0.21 \\
     55988	      &0.36 &0.30  &0.49 &0.43 &0.25 &0.14 &0.34 &0.03 &0.01 &0.29 &--0.19 &0.01 &--0.17 &	--0.23 &0.39 &0.68 \\ 
     58357        &0.12 &0.33 &0.35 &0.25 &0.26 &0.13 &0.35	 &0.07 &0.00 &0.10 &0.29 &0.27 &0.43 &0.45	&0.46 &0.35 \\
     58708        &0.21 &0.33 &0.44	 &0.33 &0.28 &0.19 &0.37 &0.10 &0.08 &0.14 &0.08 &0.03 &0.24 &--0.10 &--0.01 &0.35 \\
     58843        &0.03 &0.35 &0.20 &0.20 &0.21 &0.06 &0.25 &0.00 &--0.04 &0.09 &0.06 &0.00 &0.20 &0.00 &0.05 &0.35 \\
     59785        &0.20 &0.43 &0.43 &0.45 &0.21 &0.17 &0.19 &0.02 &0.01 &0.16 &0.08 &--0.10 &0.04 &--0.32 &--0.19 &0.37 \\
     60747        &0.01 &0.31 &0.29 &0.15 &0.24 &0.02 &0.31 &0.09 &--0.05 &0.05 &0.02 &0.04 &0.27 &0.08 &--0.07 &0.47 \\
     64920        &0.21 &0.36 &0.38 &0.29 &0.23 &0.16 &0.31 &0.03 &0.05 &0.11 &0.20 &--0.05 &0.06 &--0.10 &--0.15 &0.24 \\
     74033        &0.09 &0.36 &0.20 &0.18 &0.21 &0.12 &0.27 &0.00 &--0.03 &0.11 &0.12 &0.03 &0.34 &0.10 &0.10 &0.26 \\
    81170         &--0.16 &0.30 &0.13 &0.13 &0.41 &--0.18 &0.35 &0.05 &--0.09 &0.14 &   &0.12 &    &--0.12 &    &--0.12    \\
    87101         &--0.07 &0.53 &--0.27 &0.30 &0.35 &--0.18 &0.20 &--0.22 &0.06 &0.08 &    &--0.10 &0.05 &--0.04 &--0.20 &0.10   \\
  117720        &--0.04 &0.45 &0.37 &0.14 &0.44 &0.09 &0.51 &0.14 &--0.04 &0.07 &0.22 &0.15 &0.51 &--0.11 &    &0.39 \\
                \hline
\end{tabular}
\tablefoot{Abundances of 16 elements (Na, Mg, Al, Si, Ca, Sc, Ti, Cr, Ni, Zn, Sr, Y, Zr, Ba, La, and Eu) relative to Fe are listed in columns 2 to 17.}
\end{tiny}
\end{table*}

SME gives a typical error in abundance less than 0.01 dex due to continuum placement and line blending. However, the uncertainties in our derived abundances are dominated by the uncertainty in stellar parameters. In Sect. 3.1 the uncertainties in the stellar parameters were found to be $\sigma_{T_{\rm eff}} = 40$ K, $\sigma_{\rm{log} g}=0.06$ dex, and $\sigma_{\rm[Fe/H]}$ = 0.03 dex. The uncertainties in the elemental abundances associated with these, for two stars (HIP 18235, and HIP 64920), were calculated. The total uncertainty, shown in Table~\ref{aberr} was derived by taking the square root of the quadratic sum of the different errors, as L15 suggested. The average values of the total uncertainties for all elements are between 0.03 and 0.05 dex. Although more than two lines for the r-and s-process elements were carefully selected from the literature, there are some cases that only a single line was used to derive the elemental abundance for a given star. This could introduce more uncertainty to the abundance than what is expected. 

\begin{table*}[htbp]
\centering
\setlength{\tabcolsep}{5pt}
\begin{tiny}
\caption{Total random errors in the abundances due to the uncertainties in stellar parameters.}
\label{aberr}
\begin{tabular}{lccccccccccccccccc}
      \hline
      \hline
                      &---     &---    &---   &---  &---  &--- &---   &$\sigma$[X/Fe]    &---   &---    &---  &---  &--- &--- &--- &---  \\
      Name   &Na &Mg &Al &Si &Ca &Sc &Ti &Cr &Ni &Zn &Sr &Y &Zr &Ba &La &Eu \\
      \hline
   HIP 18235   &$\pm$0.03 &$\pm$0.05 &$\pm$0.04 &$\pm$0.03 &$\pm$0.05 &$\pm$0.03 &$\pm$0.05 &$\pm$0.04 &$\pm$0.04 &$\pm$0.03 &$\pm$0.03 &$\pm$0.05 &$\pm$0.05 &$\pm$0.03 &$\pm$0.04 &$\pm$0.04\\
   HIP 64920  &$\pm$0.04 &$\pm$0.04 &$\pm$0.03  &$\pm$0.03 &$\pm$0.05  &$\pm$0.03 &$\pm$0.05 &$\pm$0.02 &$\pm$0.02 &$\pm$0.03 &$\pm$0.02 &$\pm$0.03 &$\pm$0.03 &$\pm$0.03 &$\pm$0.03 &$\pm$0.02\\
                \hline
\end{tabular}
\end{tiny}
\end{table*}

\section{Chemical tagging}

The chemical tagging method described by \cite{2013MNRAS.428.2321M} was employed to find out whether the KFR08 stream originates from a dissolved stellar cluster. Here we give a brief summary of this method. A metric ($\delta_{C}$) was defined as
\begin{equation}
\delta_{C} = \sum_{C}^{N_{C}} \omega_{C} \frac{|A_{C}^{i}-A_{C}^{j}|}{N_{C}},
\label{ct}
\end{equation}
where $N_{C}$ is the number of measured abundances, $A_{C}^{i}$ and $A_{C}^{j}$ are individual abundance ratios of element $C$ with respect to Fe relative to solar for stars $i$ and $j$, respectively. When the element $C$ is Fe, then $A_{C}$ is the ratio of Fe to H. Here, $\omega_{C}$ is the weighting factor for an individual species. For our purposes this is set to unity as we do not know if any of the elements are more or less important. Furthermore, $\delta_{C}$ is the mean absolute difference between any two stars across all measured elements.
To turn this $\delta_{C}$ into a probability that shows how likely it is that the stars come from a dissolved cluster, we use an empirecal probability function from \cite{2013MNRAS.428.2321M} that turns the $\delta_{C}$ into a probability $P_{\delta_{C}}$.
 
Given a confidence limit $P_{lim}$, we first removed all pairs for which the probability is less than this threshold. Secondly, all the remaining candidates that make up the pairs were re-evaluated based on their $\delta_{C}$ and $P_{\delta_{C}}$. In this way we have separated the potential dissolved clusters stars from the field stars using their elemental abundances. As \cite{2013MNRAS.428.2321M} suggested, the cluster detection confidence, $P_{clus}$, is finally evaluated using the mean of $\delta_{C}$ in the tagged group. 

Two confidence limits 85\% and 68\% were used to verify whether the stream comes from a dissolved cluster or not. The 68\% threshold, which is analogous to a 1$\sigma$ detection, is the lowest meaningful probability and corresponds to a $\delta_{C}$ of 0.058 dex which is comparable with the systematic uncertainty in the abundances. As \cite{2013MNRAS.428.2321M} point out, a linked group of stars are less contaminated by other groups if a higher confidence level is used ($P_{lim}$ = 85\%).

We can not obtain abundances of [Sr/Fe], [Zr/Fe] and [La/Fe] for three stars, because the absorption lines are weak.  Thus abundances of 14 elements (Na, Mg, Al, Si, Ca, Sc, Ti, Cr, Fe, Ni, Zn, Y, Ba, and Eu) are used to identify the stream members. We found that three stars, HIP 18235, HIP 58843, and HIP 74033 could belong to one group using a limit of $P_{lim}$ = 68\%. The mean $\delta_{C}$ of the three stars is 0.044 which gives a group detection confidence of  $\sim$ 84\%. No star was tagged as a member of a  group of stars when we tested with  a higher threshold ($P_{lim}$ = 85\%).

\begin{figure*}[htbp]
\begin{center}
\includegraphics[scale=0.43]{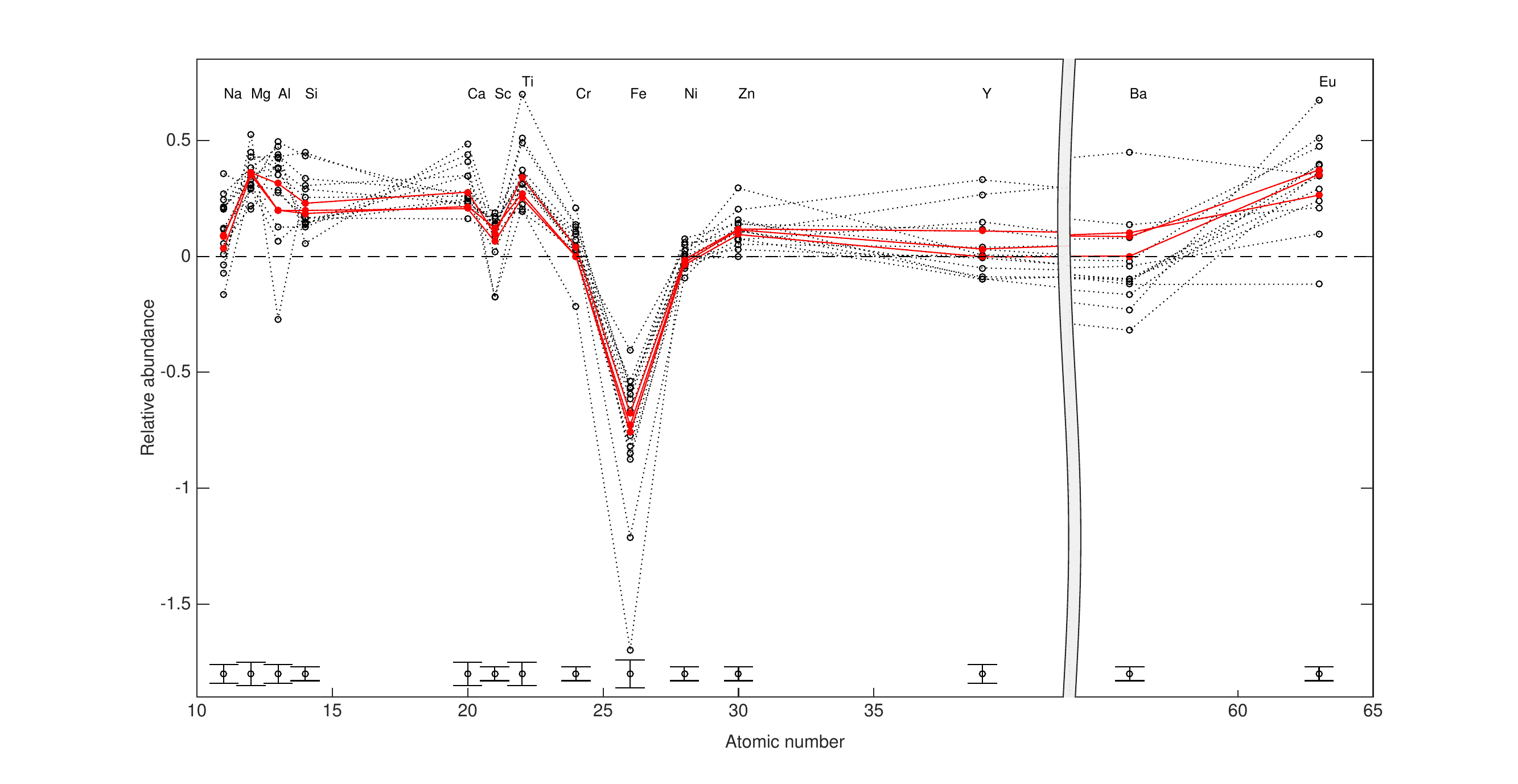}
\caption{Relative elemental abundance ratios with respect to the Sun ([X/Fe]) for all stars as a function of atomic number. When the element is Fe, the relative elemental abundance is [Fe/H]. The red dots with red lines represent the abundance patterns of the three stars tagged as cluster stars according to our chemical tagging technique (see Sect.~4), while circles with dotted lines represent the abundance patters of the other stars in our sample. The dashed line indicates the solar abundance. The gray break in the x-axis is due to the cut of blank space between Y and Ba elements.}
\label{atom}
\end{center}
\end{figure*}

In Fig.~\ref{atom}, we show the abundance patterns of the three tagged cluster stars in comparison with the other stream stars. The figure shows that the three stars (shown as red filled diamonds) almost have the same elemental abundances within the uncertainties. If all stars originated from a single star cluster, we would expect that they exhibit similar abundance ratios in all elements. This can be seen for some elements, such as Mg, Cr, and Ni. The scatter of abundance ratios among our sample stars for those elements are between 0.04 and 0.08 dex which are comparable with or relatively larger than the measurement uncertainty. However, the stars have very a large scatter ($\sigma_{\rm{[X/Fe]}}\ge 0.12$ dex) in abundance ratios in other elements, such as Na, Al, Ti, Y, Ba, and Eu. The metallicity, [Fe/H], exhibited the largest dispersion with $\sigma_{\rm{[Fe/H]}}= 0.29$. These large dispersions suggest that the stars have different birthplaces. As shown in Fig.~\ref{atom}, the star-to-star scatter of chemical composition among three tagged stars is significantly different from the other stars. This illustrates that our chemical tagging method, based on a selection of 14 elements, is efficient to isolate the cluster stars from the field stars.

As mentioned before, the stellar parameters based on the parallaxes were computed using the Procedure 2 for our sample. Then, we measured their elemental abundances using derived stellar parameters in the same way as introduced in Sect.~3. The same three stars were identified as cluster members when we followed the same chemical tagging experiment as before.

\section{Discussion} 

\subsection{Is the KFR08 a dissolved cluster?}

Recent studies on observations \citep[see][]{2007AJ....133.1161D, 2010A&A...511A..56P} and simulations \citep{2014Natur.513..523F} demonstrated that chemical compositions within star clusters are homogenous. The stream stars should have the same elemental abundances and stellar age if the KFR08 stream originated from a cluster. As mentioned in Sect.~4, the star-to-star scatter in abundances of most elements are quite large compared to the measurement uncertainty. The sample stars spread in abundances space rather than cluster together. We also found that 4 out of 16 stars (see Table~\ref{age}) are much younger than the other members. These suggest that at least a fraction of the stream members are not from a dissolved cluster.

How likely is it that our three stars found as belonging to a chemically homogenous cluster in Sect.~4 really come from a dissolved cluster and are not a chance grouping? \cite{2013MNRAS.428.2321M} predicts that half of the members in a chemically tagged group in fact are interlopers when a confidence limit of 68\% is used. Thus, it is possible that the three stars found in Sect.~4 do not belong to a chemically homogenous group of stars, i.e., a dissolved cluster. This is supported by our higher confidence level experiment that no star is linking to a group when the threshold is set to 85\%. 

In addition, if we assumed that the three stars, HIP 18235, HIP 58843, and HIP 74033, are siblings of a dissolved cluster, they should have similar RVs. However, we found that the difference in their RVs is larger than $50~\mathrm{km~s^{-1}}$ (see Table~\ref{age}). We also might expect the three stars to have similar kinematics if they were born in the cluster. When we looked at their angular momenta (see~Fig.~\ref{am}) and total velocities (see~Fig.~\ref{tom}), however, they are not more clustered than other stars in the angular momentum and velocity spaces. These further weaken their likelihood of being a dissolved stellar cluster.

The probability that the star HIP 74033 belongs to the KFR08 stream is 0.65 according to \cite{2010AstL...36...27B}. Since the location of this star in velocity space is also close to the Arcturus stream, we are not surprised that it was selected as a potential member of the Arcturus stream in \cite{2006A&A...449..533A}. Thus the nature of this star and which stream it might belong to is rather unclear.

Although we found that three stream candidates have similar elemental abundances, from this discussion we conclude that they do not belong to a dissolved cluster.

\subsection{Does the KFR08 originate from an accreted galaxy?}

\begin{figure}[htbp]
\begin{center}
\includegraphics[scale=0.4]{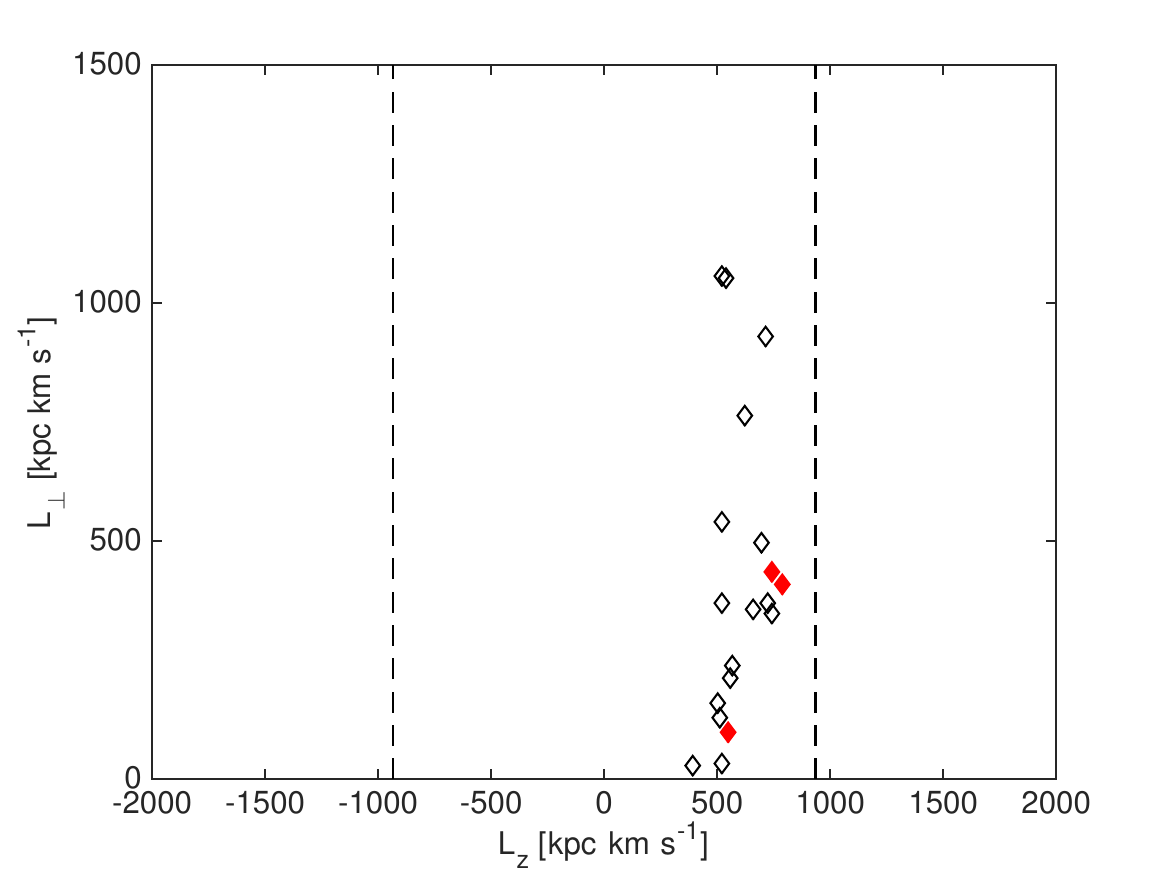}
\caption{Distribution of the stream candidates in angular momentum ($L_{\rm{z}}$ and $L_{\perp}$)-space. Note that the distance of the local standard of rest to the Galactic center is $R_{\odot}=8.5$~kpc. The coloured diamonds have the same meaning as in Fig.~\ref{abunp1}. Two dash lines are used to illustrate the angular momenta to typical halo stars.}
\label{am}
\end{center}
\end{figure}

A possible origin for the KFR08 stream is that it is part of the tidal debris from a satellite accreted long ago. Such a minor merger origin was proposed by \cite{2008ApJ...685..261K}. A significant population of stars, which lag behind the LSR by $\sim100~\rm{km~s^{-1}}$, with kinematics intermediate between the canonical thick disk and the stellar halo were interpreted as remnants of a merger by \cite{2002ApJ...574L..39G} and \cite{2006ApJ...639L..13W}. \cite{2002ApJ...574L..39G} also found more evidence that the stellar halo retains kinematic substructure indicative of minor mergers. Our sample has $V$ velocity $\simeq-152~\rm{km~s^{-1}}$ with respect to the LSR.  The stream candidates selected by \cite{2008ApJ...685..261K} from the RAVE survey look like halo stars and show quite radial orbits (high $W$ velocities). The mean $W$ velocity of our sample is $\bar{W}=-7\pm65~\rm{km~s^{-1}}$. We calculated the angular momenta of our stream members.  These are shown in Fig.~\ref{am}. We found that the stars cluster around $L_{z}=500~\rm{kpc~km~s^{-1}}$. This suggests that they might belong to the halo population. We also find that the stars have large scatter in the $L_{\perp}=(L_{x}^{2}+L_{y}^{2})^{1/2}$ component. Thus, they do not show any difference when we compare them with the distribution of normal halo stars in $L_{z}$ and $L_{\perp}$ space \citep[e.g.,][]{2007AJ....134.1579K}.

Chemically, our stars have [$\alpha$/Fe] ratios that are enhanced for any given $\alpha$-element. This is distinct from the low ratios observed in surviving dwarf satellite galaxies \citep[e.g.,][also see Sect.~5.3]{2009ARA&A..47..371T}. Thus, if the KFR08 stars were accreted, they must have come from a merger with a more massive satellite galaxy. In this case, the stars could have smaller $W$ velocities, like our observations, since they would be dragged into the plane of the Milky Way by dynamical frication \citep[e.g.,][]{2008MNRAS.389.1041R, 2014MNRAS.444..515R}.

In addition to the dynamical arguments given above, \cite{2004AJ....128.1177V} demonstrate that the thick disk cannot be comprised of the remnants from a low-mass dwarf spheroidal (dSph) galaxy, because dSph galaxies are $\alpha$-poor when compared to the Galactic stars of similar [Fe/H]. As can be seen in Fig.~\ref{abunp1} and Fig.~\ref{abunp2}, the stream candidates are $\alpha$ enhanced and [Fe/H] of the stars cover a broad range.

\subsection{Does the KFR08 have a dynamical origin?}

For the s-and r-process elements, such as Ba, La, and Eu, a large scatter in abundance ratios is seen in Fig.~\ref{abunp2}. Since they are produced in different places and time scales, this implies that our candidates might have different birth places. We collected a large sample consisting of thin and thick disk and halo stars from the literature (\citealt{2000AJ....120.1841F, 2010A&A...511L..10N, 2014A&A...562A..71B}) and show these stars together with our data in Fig.~\ref{abunp1} and \ref{abunp2}. As can be seen, the abundance patterns of our sample stars are well matched to the thick disk. It has been shown that the Eu abundance follows the Mg abundance in thin and thick disk stars while halo stars show an overabundance of Eu relative to Mg \citep{2001A&A...376..232M, 2003A&A...397..275M}. For our sample stars, the mean value $\rm{[Eu/Mg]} = -0.02 \pm 0.20$ is consistent with the trends found for the thick disk stars.

\begin{figure}[htbp]
\begin{center}
\includegraphics[scale=0.4]{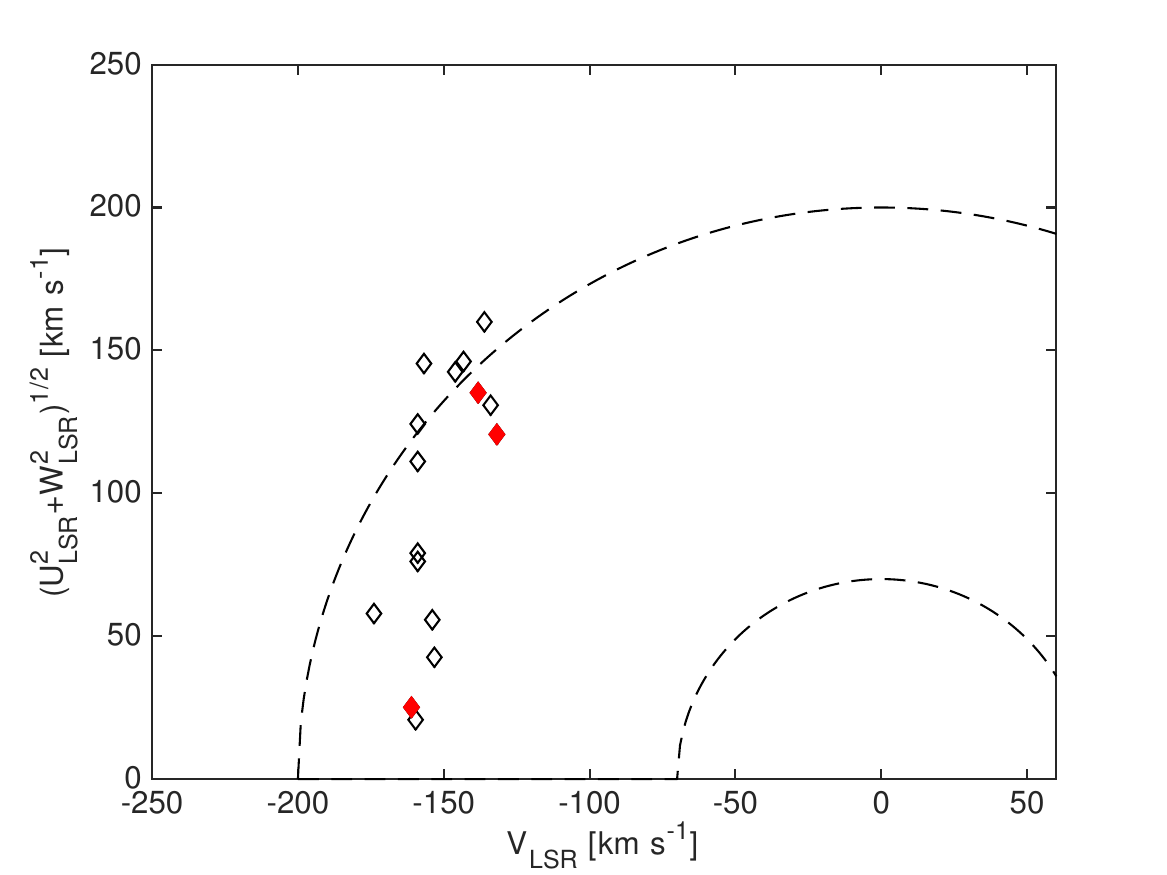}
\caption{Toomre diagram. The two circles delineate constant total space velocities of $v_{\rm{tot}}= 70$ and $200~\rm{km~s^{-1}}$, respectively. The coloured diamonds have the same meaning as in Fig.~\ref{abunp1}.}
\label{tom}
\end{center}
\end{figure}

Generally, stars with a total velocity $v_{\rm{tot}} \equiv (U^{2}+V^{2}+W^{2})^{1/2}$ greater than $\sim~70~\rm{km~s^{-1}}$, but less than $\sim~200~\rm{km~s^{-1}}$, are likely to be thick disk stars \citep[e.g.,][]{2004oee..symp..154N}. According to data presented in the Toomre diagram (Figure~\ref{tom}), the kinematics of the KFR08 stream is intermediate between the thick disk and halo populations. Combining with the chemical signature of thick disk for the stream stars, they might have once belonged to the thick disk and gained hot kinematics as a result of a satellite merger. This is consistent with the hypothesis suggested by \cite{2009MNRAS.396L..56M} that the high-velocity KFR08 stream has a dynamical origin presumably due to a strong perturbation in the Galactic disk from a merger.

As mentioned before, our sample stars have a large dispersion in $W$ velocity. The $U$ velocities also have a large scatter with $\sigma _{U} \sim~86~\rm{km~s^{-1}}$. We have shown that the KFR08 stream is older than 11 Gyr. Such old populations with high velocity dispersion could be expected if they were perturbed by massive mergers in the early universe \citep{2014ApJ...781L..20M}. It is possible that the stars were also subsequently migrated from the inner disk to their current position. However, it is unclear how much radial migration would increase the velocity dispersion of the stars \citep{2013A&A...558A...9M, 2014ApJ...794..173V}.

\section{Conclusions}

We derived the stellar parameters and elemental abundances for 16 of the KFR08 stream members identified by \cite{2010AstL...36...27B} by comparing synthetic spectra with observed spectra. The stellar ages are also obtained by fitting to isochrones. To find out whether the stream is a dissolved cluster, a chemical tagging method was used to tag the cluster members. We found that three stars have similar abundances and could belong to one group using a confidence limit of $P_{lim}=68$\%. However, further tests (see Sect. 5.1) do not support the conclusion that the three chemically similar stars come from a dissolved cluster. 

Except for a dissolved cluster origin, the stream could originate from an accreted satellite galaxy or have a dynamical origin caused by a massive merger. \cite{2008ApJ...685..261K} propose that the KFR08 stream is part of the tidal debris from a satellite accreted long ago. Although the stream stars cluster around angular momentum $L_{z} =500~\rm{kpc~km~s^{-1}}$, they have a large scatter in the $L_{\perp}$ component. This make them similar to the field halo stars in $L_{z}$ and $L_{\perp}$ space. The mean $U$ and $W$ velocities of our sample are very small with respect to the LSR. This might be inconsistent with the hypothesis that the KFR08 stream was accreted from a dwarf satellite galaxy.

We further found that the stream members are $\alpha$-enhanced. This speaks strongly against the accretion debris origin of stream if we expect the KFR08 stream progenitor to be similar to current day dSph galaxies. It has been noted that a high-mass dwarf galaxy has higher rate of star formation than the remaining satellite of the Milky Way and their metallicity can be enriched to [Fe/H] = --0.6 \citep{2001ApJ...552L.113T}. We thus can not rule out a satellite progenitor of the stream which has a substantial mass similar to the LMC or Sgr dwarf galaxy. We found that the abundance patterns of stream stars are well matched to the thick disk, especially for the [Eu/Mg]. In addition to the very old ages (> 11 Gyr), the stream stars have hotter kinematics than the canonical thick disk stars. A more likely scenario is thus that the members of the KFR08 stream have a dynamical origin due to a strong perturbation from a merger event in early universe.
   
   
\begin{acknowledgements}
The authors would like to thank Thomas Bensby for valuable stellar ages that improved the analysis of the paper. This project was completed under the GREAT -- ITN network, which is funded through the European Union Seventh Framework Programme [FP7/2007-2013] under grant agreement n$^{\rm o}$ 264895. S.F. is funded by grant No. 621-2011-5042 from The Swedish Research Council. G.R. is funded by the project grant "The New Milky Way" from the Knut and Alice Wallenberg Foundation. This research made use of the SIMBAD database, operated at the CDS, Strasbourg, France.
\end{acknowledgements}

\bibliographystyle{aa.bst} 
\bibliography{cheng-referen} 

\end{document}